\documentclass[
	groupedaddress,
	amsmath,amssymb,
	prb,
	floatfix,
	superscriptaddress,
    twocolumn,
	aps
 ]{revtex4-2}

\usepackage{graphicx}%
\usepackage{dcolumn}%
\usepackage{bm}%
\usepackage[mathlines]{lineno}%
\usepackage{xcolor}
\usepackage{soul}
\usepackage{adjustbox}
\usepackage{nth}
\usepackage{hyperref}
\usepackage{cleveref}
\usepackage[T1]{fontenc}
\usepackage{libertine}
\usepackage{nth}
\usepackage{amsmath}
\usepackage{amssymb}
\usepackage{xr}

\begin{document}

\title{Beyond special quasirandom structures: free energies from energy cumulants } 
\author{Yann L. M\"uller}
\affiliation{Laboratory of materials design and simulation (MADES), Institute of Materials, \'{E}cole Polytechnique F\'{e}d\'{e}rale de Lausanne}
\author{Anirudh Raju Natarajan}
\email{anirudh.natarajan@epfl.ch}
\affiliation{Laboratory of materials design and simulation (MADES), Institute of Materials, \'{E}cole Polytechnique F\'{e}d\'{e}rale de Lausanne}

\begin{abstract}
  Predicting the finite-temperature stability of a disordered phase requires its free energy. It is traditionally approximated by combining the energy of a special quasirandom structure (a small cell mimicking a random alloy) with the ideal configurational entropy. This approximation neglects the short-range order that develops on cooling. In this study, we show that the conventional approximation is the first term of an infinite expansion of the free energy in energy cumulants. Higher-order cumulants are estimated from energies of random arrangements, without Monte Carlo or molecular dynamics. In a nine-element refractory system, adding the second cumulant reduced free energy errors by about an order of magnitude. The expansion gives phase diagrams with order--disorder transitions, defect concentrations, and short-range order. To demonstrate its utility, we used a foundation interatomic potential to screen lithium-excess rocksalt oxides across 26 elements for synthesizable disordered phases rich in the Li$_4$ clusters needed for lithium percolation. The approach applies to any lattice and chemistry, and brings disordered phases within reach of routine screening.
\end{abstract}

\maketitle

\section{Introduction}
The free energy determines the phase stability of a material, its solubility limits, and the temperature at which a solid solution orders or decomposes. Predicting whether a candidate phase can be synthesized therefore requires its free energy at finite temperature. In multicomponent systems such as metallic high-entropy alloys and ceramics, the composition space grows combinatorially with the number of species\cite{oses2020,hsu_2024_clarifyingfour}. Computational materials design in these systems needs methods that estimate finite-temperature free energies both accurately and at low cost.

Several approximate models are widely used to meet this need. The most common, informed by \textit{ab initio} calculations, combines a temperature-independent enthalpy of the disordered phase with the ideal configurational entropy. The enthalpy is typically computed for special quasirandom structures (SQS) \cite{zunger_1990_specialquasirandom}, in which the atoms of a small supercell are arranged to mimic the correlations of a random alloy. We refer to this conventional approximation as the regular solution model. Descriptors built on it, such as the disordered enthalpy--entropy descriptor (DEED), have been proposed to rank the synthesizability of multicomponent ceramics \cite{divilov_2024_disorderedenthalpy,sarker_2018_highentropyhighhardness}. Other extensions assemble the mixing enthalpy of a chemically complex material from those of its constituent sub-systems, and have been used to screen large composition spaces \cite{wang2026, chen_2023_mapsinglephase}. Free-energy databases parameterized with the CALPHAD method provide another route to stability in high-dimensional composition spaces \cite{kaufman1970, saunders1998}. Because they are fitted largely to experiments, their accuracy is lowest in precisely the unexplored chemistries that screening studies are designed to search.

The regular solution model and its extensions neglect the short-range order (SRO) that develops as the temperature decreases, and that lowers both the enthalpy and the entropy of the disordered phase. None of these approaches predicts the SRO itself, even though many properties of disordered materials depend on it. In disordered rocksalt cathodes, for example, lithium transport requires a percolating network of lithium-rich local environments, whose abundance is set by the SRO on the cation sublattice \cite{urban_2014_configurationalspace, clement_2020_cationdisorderedrocksalt}.

More accurate atomistic routes exist, but they require a fast and accurate surrogate model and extensive sampling. An on-lattice cluster expansion (CE) can be parameterized on \textit{ab initio} data \cite{sanchez_1984_generalizedcluster,muller_2025_constructingmulticomponent,natarajan_2018_machinelearningconfigurational}. Thermodynamic properties can then be estimated from Monte Carlo (MC) sampling combined with free energy integration\cite{vanderven_2018_firstprinciplesstatistical,vandewalle_2002_selfdrivenlatticemodel}. Fitting a CE requires careful selection of training structures and clusters, and the resulting model applies only to the lattice and composition space on which it was trained. A bespoke machine-learned interatomic potential (MLIP) \cite{behler_2007_generalizedneuralnetwork,drautz_2019_atomiccluster,bartok_2010_gaussianapproximation,piersante_2026_machinelearning} can instead capture both configurational and vibrational excitations. Its parameterization is at least as demanding as that of a CE, and it requires more extensive sampling that couples MC with molecular dynamics (MD) simulations.

Fitting can be avoided altogether. The MC$^2$ method samples configurations directly with density functional theory (DFT) \cite{ghazisaeidi2021}, and foundation MLIPs trained across much of the periodic table can drive MC and MD simulations without system-specific fitting \cite{bochkarev_2024_graphatomic, lysogorskiy_2021_performantimplementation, batatia_2025_foundationmodel}. Both strategies nevertheless require a large number of energy evaluations to converge thermodynamic averages. A computational gap therefore persists. CEs sample rapidly but transfer poorly beyond their training domain, while foundation MLIPs transfer across chemistries but are too slow for the extensive MC and MD sampling that screening large composition spaces would require.

Ideally, a limited number of foundation-MLIP energy evaluations would suffice to estimate finite-temperature free energies and the SRO across arbitrary composition spaces. In this study, we develop such a method. The partition function depends on the atomic configurations only through the distribution of their energies, and this distribution is characterized by its cumulants, which include the mean, the variance, and higher-order measures of its shape. Expansions of the free energy in these cumulants are classical in statistical mechanics \cite{zwanzig1954, kubo1962}, yet they have not been used  for the configurational free energies of alloy Hamiltonians, where the SQS approximation remains the standard. We express the free energy as an expansion in these energy cumulants, in which the regular solution model is the first term. The cumulants are estimated from the energies of random atomic arrangements, without MC or MD simulation, and the resulting free energies are analytic. Phase diagrams, thermodynamic averages and, when a linear energy model such as a CE is available, the SRO can therefore be computed at little additional cost. We benchmark the expansion against MC simulations of the equiatomic alloys of a nine-component refractory system, where retaining the variance reduces the free energy error by roughly an order of magnitude relative to the regular solution model. To demonstrate its utility, we then combine the expansion with a foundation MLIP to screen lithium-excess rocksalt oxides for chemistries whose disordered phase is both synthesizable and rich in the Li$_4$ clusters needed for lithium percolation. Promising compositions identified in this way can then be verified with more accurate techniques such as coupled MC and MD sampling.

\section{Results}

\subsection{Cumulant generating functions and multicomponent free energies}
\label{sec:cumul-gener-funct}

\begin{figure*}
  \centering
  \includegraphics[width=0.9\linewidth]{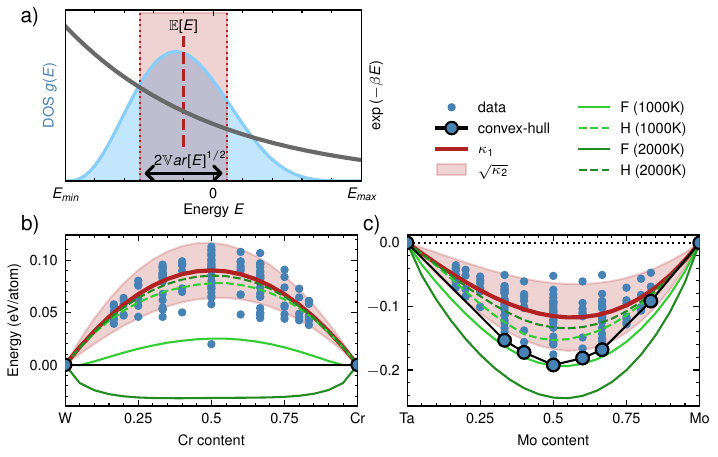}
  \caption{\textbf{Energy distributions and cumulants of disordered binary alloys.} (a) Schematic density of states $g(E)$ (blue) of a disordered phase at fixed composition. The dashed red line marks the mean $\mathbb{E}[E] = K_1$, and the shaded band spans twice the standard deviation, $2\sqrt{K_2}$. The Boltzmann factor $e^{-\beta E}$ (gray, right axis) weights $g(E)$ in the partition function of \cref{eq:mgf_beta=0}. (b and c) Energies per site of atomic arrangements in small supercells (blue circles) of (b) W$_{1-x}$Cr$_x$ and (c) Ta$_{1-x}$Mo$_x$, computed with density functional theory (DFT)\cite{muller_2025_constructingmulticomponent,muller_2026_pyecepython}. The black line is the convex hull, the red line the mean energy per site $\kappa_1$, and the shaded band $\kappa_1 \pm \sqrt{\kappa_2}$. Green lines show the free energy $F$ (solid) and enthalpy $H$ (dashed) of the binomial approximation at 1000\,K (light green) and 2000\,K (dark green). All energies are referenced to the pure elements.}
  \label{fig:DOS_cumulants}
\end{figure*}

Consider a system with discrete microstates, such as the distinct arrangements of chemical elements on a lattice. In the canonical ensemble, a microstate $\xi$ with energy $E_\xi$ occurs with the Boltzmann probability $P_\xi = \frac{e^{-\beta E_\xi}}{Z(\beta)}$, where $\beta = 1/k_B T$ is the inverse temperature and $k_B$ is Boltzmann's constant. The partition function $Z(\beta) = \sum_\xi e^{-\beta E_\xi}$ sums over all microstates accessible under the macroscopic boundary conditions, and the free energy is $F(\beta) = -\frac{1}{\beta} \ln{Z(\beta)}$.

Collecting microstates of equal energy shows that $Z$ depends on the microstates only through their energies:
\begin{equation}
  Z(\beta) = \Omega \; \mathbb{E}[e^{-\beta E}]_g
  \label{eq:mgf_beta=0}
\end{equation}
where $g(E)$ is the normalized density of states, giving the fraction of microstates with energy $E$, and $\Omega$ is the total number of microstates at a fixed composition (derivation in \cref{subsec:free_energy_approximations_validity}). At a fixed composition, the partition function is therefore proportional to the expectation value of $e^{-\beta E}$ over a probability density equal to $g(E)$ (\cref{fig:DOS_cumulants}a). The same density-of-states formulation underlies Wang--Landau MC sampling \cite{wang2001}, which estimates $g(E)$ over its full range rather than through its cumulants.

Substituting \cref{eq:mgf_beta=0} into $F(\beta)$ introduces the cumulant generating function of the energy, $\ln{\mathbb{E}[e^{-\beta E}]_g}$. Its Taylor expansion around $\beta = 0$ gives
\begin{equation}
  F(\beta) = -\frac{1}{\beta} \ln{\Omega} - \frac{1}{\beta} \sum_{n=1} \frac{(-1)^n}{n!} K_n \beta^n
  \label{eq:taylor_exp}
\end{equation}
where the expansion coefficients are the cumulants of the energy distribution,
\begin{equation}
  K_n = \left. \frac{d^n \ln{\mathbb{E}[e^{-\beta E}]_g}}{d(-\beta)^n} \right|_{\beta=0}
  \label{eq:cumulant}
\end{equation}
Here and in what follows, sums over the cumulant order $n$ run to infinity unless explicitly truncated. The first two cumulants are the mean and the variance of the energy distribution, while the third is related to its skewness. \Cref{fig:DOS_cumulants}a marks the mean and the standard deviation $\sqrt{K_2}$. The mean is the enthalpy at infinite temperature, where chemical disorder on rigid lattice sites leaves no local chemical order. The variance sets the leading high-temperature behavior of the configurational heat capacity. \Cref{eq:taylor_exp} is exact but not yet practical, because estimating every cumulant requires the full energy distribution.

Truncating the series after the first term gives an approximate free energy $\tilde{F}$,
\begin{equation}
  \label{eq:regular_solution_free_energy}
  \tilde{F}(\beta) = -\frac{1}{\beta} \ln{\Omega} + K_1
\end{equation}
in which all cumulants beyond the first are set to zero. The approximate energy distribution then collapses to a delta function at the mean energy of the composition. In the thermodynamic limit, the free energy of a binary alloy becomes $\tilde{F}(x,\beta,M) = M \kappa_{1}(x) + \frac{M}{\beta}\left(x\ln{(x)}+(1-x)\ln{(1-x)}\right)$, where $x = N/M$ is the fraction of the $M$ sites occupied by the $N$ atoms of one element and $\kappa_{1}(x) = K_{1}(x)/M$. This is the regular solution model. Its enthalpy is the temperature-independent mixing enthalpy of the disordered phase with no short-range order (referred to as a ``perfectly'' disordered phase), and its entropy is exactly the ideal configurational entropy. Because that enthalpy is usually estimated from one or a few SQSs, the SQS free energy in common use is the first term of \cref{eq:taylor_exp}. By Jensen's inequality \cite{jensen1906}, $\mathbb{E}[e^{-\beta E}]_g \geq e^{-\beta K_1}$, so \cref{eq:regular_solution_free_energy} is a rigorous upper bound on the true free energy at every temperature. An equivalent derivation from the non-negativity of the variance is given in \cref{subsec:free_energy_approximations_validity}.

Truncating \cref{eq:taylor_exp} at higher order retains more of the shape of the energy distribution. Including all terms up to quadratic in $\beta$, for instance, gives
\begin{equation}
  \label{eq:third_order_cumulant}
  \tilde{F}(\beta) = -\frac{1}{\beta} \ln{\Omega} + K_{1} -\frac{1}{2}K_{2} \beta + \frac{1}{6} K_{3} \beta^{2}
\end{equation}
Retaining $K_{2}$ and $K_{3}$ captures the temperature dependence of the enthalpy and entropy and improves on \cref{eq:regular_solution_free_energy} at high temperatures. At low temperatures, however, the truncation is not guaranteed to remain physical. Depending on the signs and magnitudes of the cumulants, the approximate free energy can diverge as $\beta \rightarrow \infty$ and the associated entropy can become negative. \Cref{eq:third_order_cumulant} can then deviate from $F(\beta)$ more than \cref{eq:regular_solution_free_energy} does, since the latter approaches $K_1$ and retains a physically meaningful entropy in the same limit.

Free energies that remain well behaved at low temperatures can instead be obtained by approximating the energy distribution itself with a probability distribution whose leading cumulants match those of the true distribution. Any such distribution has a non-negative variance, which keeps the heat capacity non-negative at every temperature. If the distribution is also bounded from below, the enthalpy remains finite as $\beta \rightarrow \infty$.
We take the per-site energy to be $\tilde{e} = \epsilon_0 + \delta X$, with $X$ a Bernoulli variable of success probability $p$. Summing over $M$ independent sites gives a shifted and scaled binomial distribution, with free energy
\begin{equation}
  \label{eq:binomial_free_energy}
  \tilde{F}(\beta) = -\frac{1}{\beta} \ln{\Omega} + M\epsilon_0 - \frac{M}{\beta} \ln{\left( 1-p + p e^{-\delta \beta} \right)}
\end{equation}
The three parameters $\epsilon_0$, $p$ and $\delta$ are fixed by matching the first three cumulants of the energy distribution (\cref{subsec:free_energy_approximations_validity}). The binomial distribution can be replaced by the distribution of any sum of $M$ independent and identically distributed random variables \cite{touchette_2009_largedeviation}. In what follows, we refer to the first-, second- and third-order truncations of \cref{eq:taylor_exp} as the regular solution, Gaussian and third-order approximations, and to \cref{eq:binomial_free_energy} as the binomial approximation. The second-order truncation is termed Gaussian because retaining only the mean and the variance is equivalent to approximating the energy distribution by a Gaussian.

\Cref{fig:DOS_cumulants}b and c compare $\kappa_1$ and $\sqrt{\kappa_2}$ with the energies per site of atomic arrangements in small supercells of the W$_{1-x}$Cr$_x$ and Ta$_{1-x}$Mo$_x$ binaries. The regular solution model fixes the enthalpy at $\kappa_1$ at every temperature, whereas the variance sets how far below $\kappa_1$ the enthalpy can fall as the temperature decreases. The energy $\kappa_1 - \sqrt{\kappa_2}$ lay within 25\,meV/atom of the convex hull over the whole composition range in Mo--Ta, and within 60\,meV/atom in W--Cr. \Cref{fig:DOS_cumulants}b and c also show the enthalpy and free energy of the binomial approximation at 1000\,K and 2000\,K for both systems.

The approximations above describe how energy levels are occupied as a function of temperature, but not the local chemical order that lowers the enthalpy and the entropy on cooling. This order is measured by the Warren--Cowley parameter \cite{cowley1965}, which for the nearest-neighbor (NN) pair of species $A$ and $B$ is $\alpha_{AB}^{NN} = 1 - \frac{\mathbb{P}_{NN}(A, B)}{x_A x_B}$. Here $\mathbb{P}_{NN}(A, B)$ is the probability that species $A$ and $B$ occupy neighboring sites, and $x_A x_B$ is that probability in a perfectly random arrangement with concentrations $x_A$ and $x_B$. Pair probabilities are linear combinations of ensemble-averaged cluster functions $\vec{\varphi}$ \cite{sanchez_1984_generalizedcluster}, so the SRO can be computed from the cluster-function averages at finite temperature. Many energy models, including on-lattice cluster expansions \cite{sanchez_1984_generalizedcluster,muller_2025_constructingmulticomponent,natarajan_2018_machinelearningconfigurational}, are linear in these cluster functions, $E = M \, \vec{\varepsilon} \cdot \vec{\varphi}$, where $\vec{\varepsilon}$ collects the interaction coefficients. %
For such models, differentiating the logarithm of the partition function with respect to the coefficients gives
\begin{equation}
    \left. \langle \vec{\varphi} \rangle \right|_\beta = -\frac{1}{\beta M} \left.\left( \frac{\partial \ln{Z}}{\partial \vec{\varepsilon}} \right)\right|_{\beta}
    \label{eq:phi}
\end{equation}
Applying \cref{eq:phi} to each approximate free energy yields an SRO estimate consistent with that approximation. The Gaussian truncation gives $\left. \langle \vec{\varphi} \rangle \right|_\beta = \vec{P}_1 - \beta \, \mathbf{P}_2 \vec{\varepsilon}$, where $\vec{P}_1$ collects the mean cluster functions and $\mathbf{P}_2$ their size-scaled covariance matrix. Both are evaluated for the perfectly disordered state at $\beta = 0$, so that $\kappa_1 = \vec{\varepsilon} \cdot \vec{P}_1$ and $\kappa_2 = \vec{\varepsilon}^{\,T} \mathbf{P}_2 \, \vec{\varepsilon}$. The regular solution model gives $\langle \vec{\varphi} \rangle = \vec{P}_1$ at every temperature, so it predicts a perfectly random arrangement even as the system cools. The third-order and binomial free energies give analogous, though more involved, expressions.

The cumulant expansion builds on earlier work by Kirkwood, Zwanzig and Kubo \cite{kirkwood_1935_statisticalmechanics,zwanzig1954,kubo1962}, and \cref{eq:phi} is closely related to the theory presented by Khachaturyan \cite{khachaturyan_1978_orderingsubstitutional}. The advantage of the present approach is that the cumulants required for the free energy, and for the SRO when the energy model is linear in the cluster functions, are estimated directly from random atomic arrangements.

\subsection{Finite-temperature properties of multicomponent refractory alloys}
\label{sec:finite-temp-prop}

We first benchmark the cumulant approximations on multicomponent refractory alloys. These alloys combine elements from groups 4, 5, and 6 of the periodic table, typically as a disordered solid solution on the body-centered cubic (BCC) lattice, and are studied for their tunable high-temperature properties \cite{natarajan_2020_crystallographythermodynamics, george_2019_highentropyalloys, george_2020_highentropy}. The finite-temperature thermodynamic properties of interest include their SRO, phase stability, and equilibrium vacancy concentrations. As references, we used the nine-component embedded cluster expansion (eCE) that we developed previously \cite{lee_2026_modelingequilibrium,muller_2026_pyecepython}, which reproduces the finite- and low-temperature phase stability of these alloys across the full Cr--Hf--Mo--Nb--Ta--Ti--V--W--Zr composition space. For the equiatomic Mo--Ta alloy, we instead used a bespoke linear CE fitted to DFT energies (\cref{sec:methods}). Canonical MC simulations driven by these models provide the reference free energies against which the approximations are compared.

\begin{figure}
    \centering
    \includegraphics[width=0.85\linewidth]{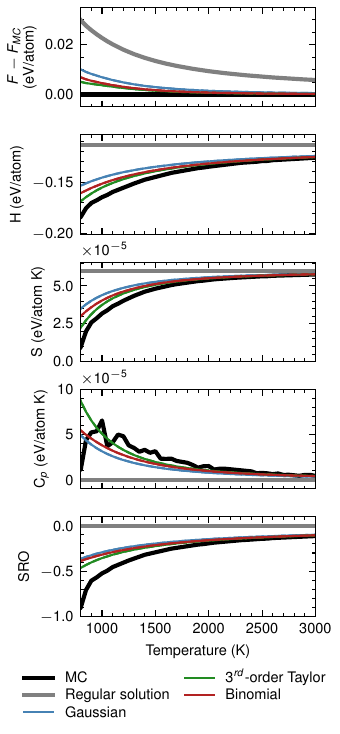}
    \caption{\textbf{Thermodynamic properties of the equiatomic Mo--Ta BCC alloy.} Panels a to e run from top to bottom. (a) Free energy of each approximation relative to the canonical MC free energy, $F - F_{\text{MC}}$. (b) Enthalpy $H$. (c) Configurational entropy $S$. (d) Heat capacity $C_p$. (e) Nearest-neighbor Mo--Ta Warren--Cowley parameter. Black lines show canonical MC results driven by a binary cluster expansion, with ensemble-averaged energies converged to 1\,meV/atom. Colored lines show the regular solution (gray), Gaussian (blue), third-order (green) and binomial (red) approximations. All quantities are per atom.}
    \label{fig:Mo-Ta_compare}
\end{figure}

\Cref{fig:Mo-Ta_compare}a compares the free energy of the equiatomic Mo--Ta BCC alloy from canonical MC with the four cumulant approximations. At high temperatures, all approximations beyond the regular solution model agreed closely with the MC free energy. By retaining the variance, the Gaussian approximation improved substantially on the regular solution model, while the third-order and binomial approximations gave smaller gains.

The enthalpy, entropy and heat capacity separated the approximations more clearly (\cref{fig:Mo-Ta_compare}b--d). The regular solution model predicted a temperature-independent enthalpy, whereas the Gaussian approximation reproduced the overall temperature dependence of the MC enthalpy. The third-order and binomial approximations improved on it slightly. The heat capacity of the regular solution model vanishes identically. The other three approximations captured the rise of the MC heat capacity on cooling, but not its peak near 1000\,K. The entropy showed the largest discrepancies below the temperature of the heat-capacity peak, and hence below the order--disorder transition of the MC simulation cell.

Both the enthalpy and the entropy decrease monotonically with increasing $\beta$, because their derivatives are proportional to the variance $\langle E^2 \rangle - \langle E \rangle^2$, and the entropy must remain non-negative. These two conditions set the range of temperatures over which each truncation remains valid (\cref{subsec:free_energy_approximations_validity}). For the second-order expansion, the entropy condition gives a critical value $\beta_c = \sqrt{2\ln\Omega/K_2}$. For the third-order expansion, the applicable condition depends on the sign of $\kappa_3$. The resulting critical temperatures lay below the order--disorder transition temperature observed in the MC simulations, so each truncation broke down only where the disordered phase was already unstable. Between 800\,K and 3000\,K, the third-order and binomial approximations deviated from the MC free energy of the equiatomic Mo--Ta alloy by at most 5\,meV/atom, and the Gaussian approximation by at most 10\,meV/atom, against up to 30\,meV/atom for the regular solution model.

\Cref{fig:Mo-Ta_compare}e shows the nearest-neighbor Mo--Ta SRO predicted by each approximation, together with the canonical MC values. The regular solution model predicted zero SRO at all temperatures, because it assumes a perfectly random solid solution. The Gaussian, third-order and binomial approximations all predicted a negative Warren--Cowley parameter at every temperature. A negative value indicates a preference for unlike nearest neighbors, and hence a tendency towards Mo--Ta ordering. The third-order approximation matched the MC values most closely.

\begin{figure}
    \centering
    \includegraphics[width=0.8\linewidth]{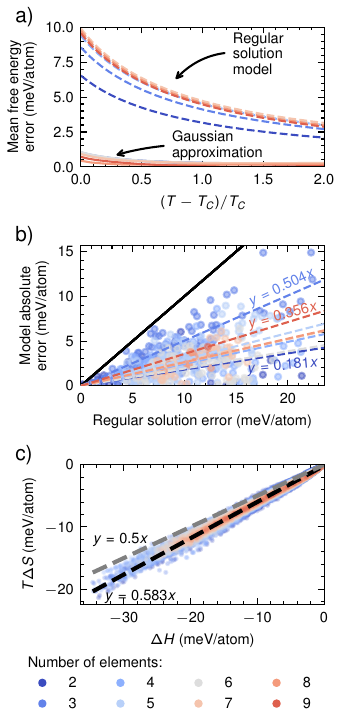}
    \caption{\textbf{Accuracy of the cumulant approximations across the nine-component Cr--Hf--Mo--Nb--Ta--Ti--V--W--Zr system.} (a) Mean absolute error of the free energy relative to canonical MC for the regular solution model (dashed) and the Gaussian approximation (solid), as a function of the reduced temperature $(T - T_c)/T_c$, where $T_c$ is an approximate order--disorder transition temperature of each alloy. Each curve averages over the equiatomic alloys with a given number of elements. (b) Absolute free energy error at $T_c$ of the binomial approximation against that of the regular solution model, both built from the energies of random 128-atom cells. The binomial cumulants were estimated from a pair cluster expansion fitted to those energies. Each point is one equiatomic alloy. Dashed lines are linear fits for each number of elements, and the solid black line marks equal errors. (c) Deviation of the entropy term $T\Delta S$ against that of the enthalpy $\Delta H$ from their regular solution values, for every equiatomic alloy at several temperatures above $T_c$. The black dashed line is the best fit, $y = 0.583x$, and the gray dashed line is the leading-order prediction $y = x/2$ of \cref{eq:deltaH_deltaS}. Colors give the number of elements in each alloy.}
    \label{fig:multicomponent}
\end{figure}

After the single binary alloy, we extended the benchmark to all 502 equiatomic alloys of two to nine elements in the nine-component system. \Cref{fig:multicomponent}a compares the mean absolute free energy error of the regular solution model with that of the Gaussian approximation, as a function of the reduced temperature $(T - T_c)/T_c$ above the order--disorder transition. For alloys with the same number of elements, the largest average regular solution error ranged from 6.5 to 10\,meV/atom, while the average error of the Gaussian approximation stayed below 1\,meV/atom. Retaining the variance therefore reduced the error by roughly an order of magnitude at every reduced temperature shown.

We next tested the approximations in a setting similar to that in which disordered free energies are usually estimated, where SQSs of roughly 100 atoms characterize the disordered phase. We mimicked this setting with random decorations of 128-atom BCC cells containing an equiatomic mixture of two to nine elements. Their energies were used both to build a regular solution model and to parameterize a pair cluster expansion, from which the higher cumulants were estimated (\cref{sec:methods}). \Cref{fig:multicomponent}b compares the free energy error of the binomial approximation with that of the regular solution model at the critical temperature. A linear fit for each number of elements gave slopes between 0.18 and 0.50, so the binomial error was on average 18 to 50\% of the regular solution error. The reduction was largest for binary alloys and smallest for ternary alloys.

The regular solution model incurs errors in both the enthalpy and the entropy. As the temperature decreases, the enthalpy of the alloy falls towards the ground-state energy, and the entropy drops below its ideal value as SRO develops. \Cref{fig:multicomponent}c compares the two errors, $\Delta H$ and $T\Delta S$, for every equiatomic alloy of the nine-component system at several temperatures above the order--disorder transition, evaluated against canonical MC. The two errors were strongly correlated, and their ratio stayed close to $1/2$ regardless of the chemical system, with a best-fit slope of 0.583.

\subsection{Screening disordered Li-excess rocksalt oxides}
\label{sec:screening}

Disordered rocksalt oxides are candidate lithium-ion cathodes in which lithium and transition metals share a single cation sublattice. Lithium transport through such a structure requires a percolating network of lithium-rich environments, and alloying the cation sublattice with several elements can promote the aggregation of lithium into the Li$_4$ clusters that enable percolation \cite{urban_2014_configurationalspace, clement_2020_cationdisorderedrocksalt, liu_2025_tailoredordering,liu_2026_shortrangeorder}. A useful candidate material therefore combines a high probability of forming Li$_4$ clusters with a low critical temperature for decomposition of the disordered phase, so that the disordered state remains stable down to accessible synthesis temperatures. We screened oxides of composition Li$_{1.2}$A$_{0.2}$B$_{0.2}$C$_{0.2}$D$_{0.2}$O$_2$. The cations A, B, C and D were drawn, with repetition allowed, from a palette of 26 elements: Na, Mg, Al, Ca, Ti, V, Cr, Mn, Fe, Co, Ni, Cu, Zn, Ga, Ge, Zr, Nb, Mo, Ru, Rh, Pd, Ag, Cd, Hf, Ta and W. Applying the criteria of Liu \textit{et al.} \cite{liu_2025_tailoredordering}, we retained only the systems that maintain charge balance and allow the extraction of at least 0.5 Li per formula unit.

Each oxide was treated as a pseudo-binary system on the cation sublattice. The four alloying elements were represented by a single mean-field cation M, so that the cation sublattice carries only Li and M, while the anion sublattice remains fully occupied by oxygen. For each oxide, a pseudo-binary cluster expansion was parameterized on the energies of relaxed random arrangements evaluated with the graph atomic cluster expansion (GRACE) foundation MLIP \cite{bochkarev_2024_graphatomic, lysogorskiy_2021_performantimplementation} (\cref{subsec:screening_methods}). The multivariate cumulants were then estimated from a single set of random decorations, without any thermodynamic sampling. The probability of forming a Li$_4$ cluster was computed from the binomial approximation of \cref{eq:binomial_free_energy} combined with \cref{eq:phi}.

\begin{figure*}
    \centering
    \includegraphics[width=0.9\linewidth]{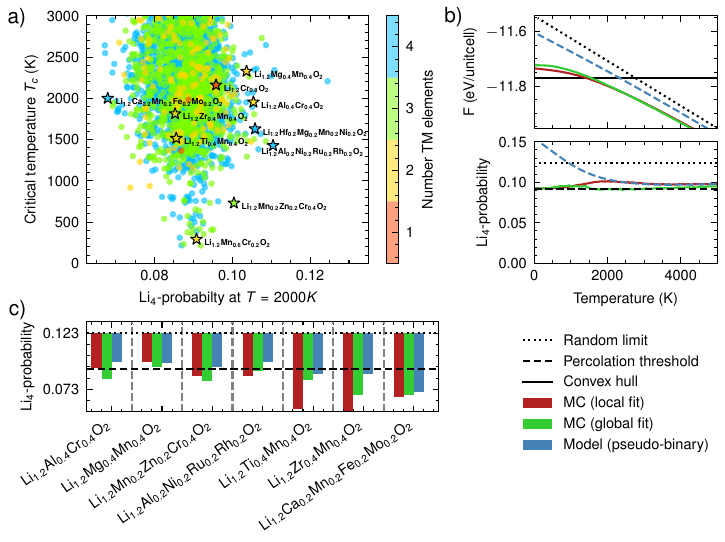}
    \caption{\textbf{Screening of Li-excess disordered rocksalt oxides.} (a) Critical temperature $T_c$ of the disordered phase against its Li$_4$-cluster probability at 2000\,K, from the pseudo-binary cumulant model, for every Li$_{1.2}$A$_{0.2}$B$_{0.2}$C$_{0.2}$D$_{0.2}$O$_2$ oxide that passed the charge-balance and Li-extraction filters. Colors give the number of distinct elements on the M sites, and stars mark the labeled oxides. (b) Free energy of the disordered phase (upper) and Li$_4$-cluster probability (lower) of Li$_{1.2}$Mg$_{0.4}$Mn$_{0.4}$O$_2$ as a function of temperature. The upper panel compares the regular solution model (black dotted), the pseudo-binary binomial model (blue dashed) and canonical MC (red and green) with the convex hull of the competing ground states (black solid). The lower panel marks the random limit (dotted) and the percolation threshold (dashed). (c) Li$_4$-cluster probabilities of seven representative oxides from the pseudo-binary model (blue) and from canonical MC (red and green). In (b) and (c), MC used a cluster expansion fitted either at the composition of the oxide (local fit, red) or across the full Li--M composition space of the same chemistry (global fit, green). All energies were computed with the GRACE foundation MLIP.}
    \label{fig:screening}
\end{figure*}

\Cref{fig:screening}a summarizes the screen across every oxide that passed the composition filters, plotting the critical temperature of each against its Li$_4$-cluster probability estimated at 2000\,K, with the color giving the number of distinct elements on the M sites. Useful cathodes lie in the lower right, at low $T_c$ and high Li$_4$-cluster probability. Chemistries with Li$_4$-cluster probabilities above the percolation threshold of about 9\% included Li$_{1.2}$Mg$_{0.4}$Mn$_{0.4}$O$_2$, Li$_{1.2}$Mn$_{0.2}$Zn$_{0.2}$Cr$_{0.4}$O$_2$ and Li$_{1.2}$Al$_{0.2}$Ni$_{0.2}$Ru$_{0.2}$Rh$_{0.2}$O$_2$, at 10 to 11\% at 2000\,K. Several oxides containing Mn and Cr also combined critical temperatures below 1000\,K with Li$_4$-cluster probabilities above 9\%, and are the most promising candidates identified by the screen. 

\Cref{fig:screening}b shows the temperature dependence of both screening criteria for Li$_{1.2}$Mg$_{0.4}$Mn$_{0.4}$O$_2$. In the binomial approximation, the free energy of the disordered phase falls below the convex hull of its competing ground states above 2329\,K. The ground-state energies were taken from the Materials Project \cite{jain2013}. For comparison, we performed canonical MC with two cluster expansions fitted to GRACE energies. The ``local'' expansion was fitted to Li--M orderings at the composition of the oxide, and the ``global'' expansion to Li--M orderings across the full composition space of the same chemistry. Above the critical temperature, the three estimates of the Li$_4$-cluster probability agreed at about 10\%, below the random limit of 12.3\%. Below about 1500\,K, the pseudo-binary estimate rose to 15\% at 0\,K, whereas both MC estimates stayed between 9 and 10\%. The regular solution model overestimated the MC critical temperature by nearly 1200\,K, and the cumulant model reduced this error to about 750\,K. The remaining error could be reduced further by retaining cumulants beyond the third, by estimating the cumulants from a larger set of random arrangements, or by treating each cation as a separate species instead of using the pseudo-binary approximation.

\Cref{fig:screening}c compares the Li$_4$-cluster probabilities of seven representative oxides from the pseudo-binary model with the local and global MC references. The pseudo-binary model predicted a higher Li$_4$-cluster probability than both MC references for every oxide, by up to about 3 percentage points, and placed four of the seven oxides on the same side of the percolation threshold as both references. Both MC references gave a higher Li$_4$-cluster probability for Li$_{1.2}$Ti$_{0.4}$Mn$_{0.4}$O$_2$ than for Li$_{1.2}$Zr$_{0.4}$Mn$_{0.4}$O$_2$. The cumulant-based estimate of SRO predicted the same ordering, with a smaller difference between the two oxides than either MC reference.

\section{Discussion}

The results above establish cumulant generating functions as a convenient route to the thermodynamic properties of disordered phases at elevated temperatures. Approximations that retain the variance and higher cumulants of the energy distribution tracked the MC free energy of Mo--Ta to within 5 to 10\,meV/atom, against up to 30\,meV/atom for the regular solution model (\cref{fig:Mo-Ta_compare}). Across the equiatomic alloys of the nine-component BCC system, retaining the variance reduced the free energy error by roughly an order of magnitude (\cref{fig:multicomponent}a). The improvement persisted when the cumulants were estimated from small disordered cells (\cref{fig:multicomponent}b). The same approximations also predicted the sign of the Mo--Ta SRO, which the regular solution model sets to zero at every temperature. The energy $\kappa_1 - \sqrt{\kappa_2}$ lies within 25\,meV/atom of the convex hull in Mo--Ta (\cref{fig:DOS_cumulants}c), which suggests that the variance alone already accounts for most of the enthalpy reduction available to the disordered phase on cooling.

The formalism also clarifies the relationship to existing thermodynamic models. Models that combine an SQS-derived disordered enthalpy with the ideal configurational entropy \cite{zunger_1990_specialquasirandom} correspond exactly to the first-order truncation of \cref{eq:taylor_exp} given in \cref{eq:regular_solution_free_energy}. The cumulant expansion therefore embeds these models within a systematic hierarchy, in which each retained cumulant adds a physically interpretable correction, beginning with the variance that controls the heat capacity of the disordered phase. In practice the disordered enthalpy is already averaged over the energies of several SQSs, while the variance of those same energies is discarded, even though it supplies the leading correction to the free energy.

The expansion further explains why the regular solution model is more accurate than the errors in its enthalpy or entropy alone would suggest (\cref{fig:Mo-Ta_compare}). As the temperature decreases, the enthalpy falls towards the ground-state energy and the entropy drops below its ideal value as SRO develops, and the two errors partially cancel. Writing the differences between the true enthalpy and entropy per site, $h$ and $s$, and the regular solution values gives
\begin{equation}
    \begin{split}
        \Delta h &= h - h_{\text{dis}} = \sum_{n=2} \frac{(-1)^{n-1}}{(n-1)!} \kappa_n \beta^{n-1} \\
        T\Delta s &= T(s - s_{\text{ideal}}) = \sum_{n=2} \left(1-\frac{1}{n}\right)\frac{(-1)^{n-1}}{(n-1)!} \kappa_n \beta^{n-1}
    \end{split}
    \label{eq:deltaH_deltaS}
\end{equation}
To leading order in $\beta$, the ratio $T\Delta s / \Delta h$ is exactly $1/2$, independent of both the cumulants and the temperature, which accounts for the near-constant ratio across chemical systems in \cref{fig:multicomponent}c. For a Gaussian density of states all higher cumulants vanish and the ratio is $1/2$ at every temperature. The fitted slope of 0.583 therefore reflects the higher-order terms in \cref{eq:deltaH_deltaS}, which depend on both the cumulants and the temperature.

\begin{figure*}
    \centering
    \includegraphics[width=0.9\linewidth]{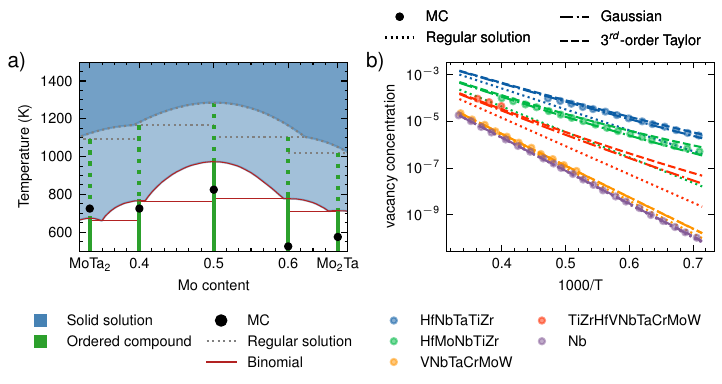}
    \caption{\textbf{Phase diagram and vacancy concentrations from the cumulant expansion.} (a) Temperature--composition phase diagram of the BCC Mo--Ta alloy between MoTa$_2$ and Mo$_2$Ta, with the ordered compounds (green) treated as line compounds. The solid red and dashed gray lines are the phase boundaries computed with the binomial approximation and the regular solution model, respectively. The solid solution is stable in the dark blue region according to both approximations, and in the light blue region according to the binomial approximation only. Solid green lines mark where each ordered compound is stable in the binomial approximation, and dotted green lines extend this range to where it is stable in the regular solution model. Black circles are order--disorder transition temperatures from canonical MC combined with free energy integration. (b) Equilibrium vacancy concentration against $1000/T$ for pure Nb and four equiatomic refractory alloys, from the canonical MC simulations of Lee \textit{et al.} \cite{lee_2026_modelingequilibrium} (circles) and from the regular solution (dotted), Gaussian (dash-dotted) and third-order (dashed) approximations of \cref{eq:vacancy_concentration}. Colors identify the alloys.}
    \label{fig:extensions}
\end{figure*}

Beyond the free energy of a single disordered phase, the cumulant expansion can be used to compute finite-temperature phase diagrams and point-defect properties. \Cref{eq:taylor_exp} factorizes into a composition-dependent part, entering through the cumulants $\kappa_n(x)$, and a temperature-dependent part, entering through powers of $\beta$. This separation of variables parallels the structure of CALPHAD free-energy models, and it means that once the composition dependence of the cumulants is known, phase boundaries can be located analytically. \Cref{fig:extensions}a shows a full temperature--composition phase diagram obtained this way, including the ordered phases. At the equiatomic composition, the order--disorder transition temperature of the regular solution model differs from that of MC by nearly 50\%, while the binomial approximation of \cref{eq:binomial_free_energy} reduces this error to about 10\%. At other compositions, the binomial error is in places close to zero, and its largest value is half that of the regular solution model. The binomial approximation therefore gives the more accurate transition temperature at every composition. The same separation gives access to point-defect properties. Because the cumulants vary linearly with the vacancy concentration $x_v$ in the dilute limit, $\kappa_n(x_v) \approx \kappa_n^\circ + x_v \, \delta \kappa_n$, minimizing the free energy with respect to the number of vacancies gives
\begin{equation}
    x_v \approx \exp{\left( \sum_{n=1} \frac{(-1)^n \beta^n}{n!} (\kappa_n^\circ + \delta \kappa_n) \right)}
    \label{eq:vacancy_concentration}
\end{equation}
\Cref{fig:extensions}b compares the vacancy concentrations of the regular solution model and the cumulant expansion with canonical MC. The regular solution model underestimates the equilibrium vacancy concentration by almost an order of magnitude, and retaining the second cumulant removes most of that discrepancy.

The screening of Li-excess rocksalt oxides illustrates how cheaply large composition spaces can be covered, while its predictions rest on assumptions beyond the free energy model itself. Treating the senary oxide as a pseudo-binary system introduces an error of its own, which \cref{fig:screening}c quantifies against canonical MC, and the estimated cumulants inherit the accuracy of the GRACE foundation MLIP. The screen should therefore be viewed as a ranking of candidate compositions rather than as a quantitative prediction of either the critical temperature or the Li$_4$-cluster probability. Experimentally, Li$_{1.2}$Ti$_{0.4}$Mn$_{0.4}$O$_2$ is among the best-performing disordered rocksalt cathodes, while Li$_{1.2}$Zr$_{0.4}$Mn$_{0.4}$O$_2$ shows poorer lithium transport, which has been attributed to a lower Li$_4$-cluster probability \cite{ji_2019_hiddenstructural}. The pseudo-binary model and both MC references ranked these two oxides in the same order (\cref{fig:screening}c). Because the pseudo-binary model overestimates the Li$_4$-cluster probability, candidates near the percolation threshold require verification with more accurate methods.

While the cumulant-based method has clear advantages, several limitations remain. The truncated expansions lose validity below a critical temperature at which the variance or the entropy would become negative. In the systems studied, this breakdown occurred only where the disordered phase was already unstable with respect to ordering or decomposition. However, there is no guarantee that the two temperatures order this way in a system with a strongly skewed density of states. The binomial approximation mitigates the low-temperature pathology, at the cost of slightly reduced accuracy at elevated temperatures when a vanishing zero-temperature entropy is enforced. The accuracy of the estimated cumulants is inherited from the underlying energy model, and the SRO is accessible only when that model is linear in descriptors of the atomic configuration. The free energies in this study are configurational, and vibrational contributions are not included.

Taken together, these results provide a route to the finite-temperature properties of disordered phases that avoids extensive thermodynamic sampling and free energy integration. A limited number of energy evaluations of random atomic arrangements, obtained for instance from a foundation MLIP, suffices to construct analytic free energy models across arbitrary composition spaces. The resulting estimates of phase stability and SRO can narrow a large composition space to a small set of candidates, for which coupled MC and MD sampling becomes affordable. The cumulant-based free energy model presented in this study unlocks a rigorous path towards the high-throughput screening of finite-temperature properties of multicomponent alloys and ceramics.

\section{Materials and Methods}
\label{sec:methods}

\subsection{Derivations and parameterization of the free energy approximations}
\label{subsec:free_energy_approximations_validity}

\textit{Partition function and upper bound.} Collecting microstates of equal energy gives \cref{eq:mgf_beta=0},
\begin{equation}
    \begin{split}
        Z(\beta) &= \sum_\xi \int_{-\infty}^{\infty} e^{-\beta E} \delta(E - E_\xi) \,dE \\
        &= \Omega \int_{-\infty}^{\infty} e^{-\beta E} g(E) \,dE = \Omega \; \mathbb{E}[e^{-\beta E}]_g
    \end{split}
\end{equation}
The variance of the energy is non-negative at every temperature, and truncating \cref{eq:taylor_exp} at the first cumulant sets it to zero, its lowest permitted value. Integrating this inequality twice over $\beta$ makes \cref{eq:regular_solution_free_energy} an upper bound on the true free energy:
\begin{equation}
    \begin{split}
    \left. \left[ \langle E^2 \rangle - \langle E \rangle^2 \right] \right|_\beta &= \left. \frac{d^2 \ln{Z}}{d\beta^2} \right|_\beta = -\left. \frac{d \langle E \rangle }{d\beta} \right|_\beta \geq 0 \\
    \left. \langle E \rangle \right|_\beta - K_1 &= \int_0^{\beta} -\left. \left[ \langle E^2 \rangle - \langle E \rangle^2 \right] \right|_x dx \leq 0 \\
    \ln{Z(\beta)} - \ln{\Omega} &= \int_0^{\beta} -\left. \langle E \rangle \right|_x dx \geq -K_1 \beta \\
    F(\beta) &=-\frac{1}{\beta} \ln{Z(\beta)} \leq -\frac{1}{\beta} \ln{\Omega} + K_1
    \end{split}
    \label{eq:upper_bound}
\end{equation}

\textit{Binomial parameters.} We fixed the three parameters of the binomial free energy of \cref{eq:binomial_free_energy} by matching the first three cumulants per site,
\begin{equation}
\begin{split}
    \delta &= \sqrt{4\kappa_2 + \left(\frac{\kappa_3}{\kappa_2}\right)^2} \\
    p &= \frac{1}{2} \left( 1 - \frac{\kappa_3}{\kappa_2 \delta} \right) \\
    \epsilon_0 &= \kappa_1 - p \delta
\end{split}
\label{eq:bernoulli_free_energy}
\end{equation}
An alternative parameterization matches only the first two cumulants and requires the entropy to vanish at zero temperature. It reduces the flexibility of the model and gives slightly worse agreement at elevated temperatures, while ensuring physically meaningful behavior at low temperatures.

\textit{Range of validity.} Each truncation remains valid only where the variance and the entropy of the approximate model stay non-negative,
\begin{subequations}
\begin{align}
     \left. \left[\langle E^2 \rangle - \langle E \rangle^2 \right] \right|_\beta = \left. \frac{d^2 \ln{Z}}{d\beta^2} \right|_\beta &\geq 0 \label{eq:criteria_variance} \\
     S(\beta) = k_B \left(\ln{Z(\beta)} - \beta \left. \frac{d\ln{Z}}{d\beta} \right|_\beta \right) &\geq 0 \label{eq:criteria_entropy}
\end{align}
\end{subequations}
For the second-order expansion, \cref{eq:criteria_entropy} gives the critical value $\beta_c = \sqrt{2\ln\Omega/K_2}$. For the third-order expansion, the applicable criterion depends on the sign of $\kappa_3$: we applied \cref{eq:criteria_variance} when the density of states is positively skewed and \cref{eq:criteria_entropy} otherwise. Below the corresponding critical temperature we held the enthalpy constant and set the entropy to zero, to maintain numerical stability.

\subsection{Atomistic models}

We described the finite-temperature thermodynamics of the BCC refractory alloys with the nine-component embedded cluster expansion (eCE) developed previously \cite{muller_2026_pyecepython}, parameterized across the full Cr--Hf--Mo--Nb--Ta--Ti--V--W--Zr composition space. Equilibrium vacancy concentrations were evaluated with the ten-component eCE of Lee \textit{et al.} \cite{lee_2026_modelingequilibrium}, in which the vacancy is treated as an additional species on the BCC lattice.

We fitted two smaller models in this work. For the SRO of the equiatomic Mo--Ta alloy, we fitted a cluster expansion of 9 pairs, 9 triplets and 1 quadruplet to the DFT energies of 137 decorations of small cells from a previous study \cite{muller_2025_constructingmulticomponent}. The decorations are all symmetrically distinct arrangements of up to 6 atoms, enumerated as described in \cite{hart_2008_algorithmgenerating,hart_2009_generatingderivative}. The resulting CE had a training root-mean-square error (RMSE) of 3.7\,meV/atom and a fivefold cross-validation error of 4.1\,meV/atom, at an optimal regularization strength of 0.0001.

For the benchmark of \cref{fig:multicomponent}b, we fitted a cluster expansion of 2 pairs for each alloy to the energies of $3 N_\varphi$ decorations of 128-atom BCC cells, where $N_\varphi$ is the number of cluster functions of the alloy. The decorations were selected from 1000 random arrangements at the given composition by choosing those farthest apart from each other, so that the training data sample the space of cluster functions as widely and homogeneously as possible. The selected decorations are SQS-like in size but contain SRO, which improves the parameterization.

\subsection{Screening of Li-excess rocksalt oxides}
\label{subsec:screening_methods}

We enumerated all oxides of composition Li$_{1.2}$A$_{0.2}$B$_{0.2}$C$_{0.2}$D$_{0.2}$O$_2$, with A, B, C and D drawn with repetition from the 26-element palette listed in \cref{sec:screening}. Applying the criteria of Liu \textit{et al.} \cite{liu_2025_tailoredordering}, we retained only systems that maintain charge balance and allow the extraction of at least 0.5 Li per formula unit. Each oxide was represented in a 216-atom supercell of composition Li$_{64}$A$_{11}$B$_{11}$C$_{11}$D$_{11}$O$_{108}$, which approximates the Li$_{1.2}$A$_{0.2}$B$_{0.2}$C$_{0.2}$D$_{0.2}$O$_2$ stoichiometry. For each oxide, a pseudo-binary cluster expansion of 5 pairs, 7 triplets and 1 quadruplet on the cation sublattice, in which the four alloying elements are represented by a single mean-field cation M, was parameterized on the energies of 100 relaxed random arrangements evaluated with GRACE-2L-OMAT-large-ft-AM\cite{bochkarev_2024_graphatomic,lysogorskiy_2021_performantimplementation}. 
The structures were relaxed with the ``good old quasi-Newton'' optimizer of ASE until the maximum force fell below 0.05\,eV/\AA.
The multivariate cumulants were then estimated from a single set of random decorations, and the Li$_4$-cluster probability was computed as a linear combination of the ensemble-averaged cluster functions obtained from \cref{eq:binomial_free_energy} combined with \cref{eq:phi}. Critical temperatures were obtained by comparing the free energy of the disordered phase with the convex hull of its competing ground states, whose energies were taken from the Materials Project \cite{jain2013}.

For the validation of \cref{fig:screening}c, canonical MC was performed in a $6 \times 6 \times 6$ conventional FCC supercell of the cation sublattice with two cluster expansions per oxide. The local fit was parameterized on 1000 randomly generated configurations at the composition of the screened oxide only, while the global fit was parameterized on enumerated Li--M orderings of up to 6 unit cells spanning the full composition space of the same chemistry. The RMSEs of the local and global fits were below 20\,meV and 50\,meV per unit cell, respectively.

\subsection{Sampling the energy distribution and estimating cumulants}
\label{sec:cumulant}

We estimated the cumulants from the energies of 10$^{6}$ random arrangements at fixed composition in $10 \times 10 \times 10$ conventional BCC or $6 \times 6 \times 6$ conventional FCC supercells, whose energies are distributed according to $g(E)$. \Cref{fig:multicomponent}a used 2000 arrangements instead. Univariate cumulants were computed as k-statistics with \textit{SciPy}, and the multivariate cumulants required by \cref{eq:phi} from their relation to the central moments.

\subsection{Monte Carlo simulations}

Reference free energies were obtained from canonical MC simulations in a $10 \times 10 \times 10$ conventional BCC supercell or $6 \times 6 \times 6$ conventional FCC supercell driven by the cluster expansion models, combined with free energy integration. At each temperature, at least 100 uncorrelated samples were drawn, and the average energy was converged to within 1\,meV/atom.

\subsection{Computing phase diagrams with the cumulant approximation}
Computing the BCC Mo--Ta phase diagram of \cref{fig:extensions}a requires the free energy of the disordered phase at all compositions and at each temperature of interest.
We estimated the cumulants at 21 equally spaced compositions from pure Ta ($x=0$) to pure Mo ($x=1$), as described in \cref{sec:cumulant}, and interpolated them at intermediate compositions with the piecewise cubic interpolator of \textit{SciPy}.

The solid solution competes with the ordered structures on the convex hull of \cref{fig:DOS_cumulants}c, which we treated as line compounds. At each temperature, we computed the convex hull of the binomial free energy of the solid solution (\cref{eq:binomial_free_energy}), evaluated at 1001 equally spaced compositions, together with the free energies of the ordered phases, using the \textit{Qhull} algorithm. The regular solution boundaries were computed in the same way from \cref{eq:regular_solution_free_energy}.

\subsection{Vacancy thermodynamics}

To obtain the dilute-limit parameters $\kappa_n^\circ$ and $\delta \kappa_n$ of \cref{eq:vacancy_concentration}, we estimated the cumulants at vacancy concentrations of 0, 0.05, 0.1 and 0.5\%, corresponding to 0, 1, 2 and 10 vacancies in the $10 \times 10 \times 10$ BCC supercell (\cref{sec:cumulant}), and fitted their linear dependence on $x_v$ by least squares.

\section{Acknowledgments}
ARN is grateful for support through the framework of the EUROfusion Consortium, partially funded by the European Union via the Euratom Research and Training Programme (grant agreement no. 101052200--EUROfusion). Views and opinions expressed are however those of the author(s) only and do not necessarily reflect those of the European Union or the European Commission. Neither the European Union nor the European Commission can be held responsible for them.

\end{document}